\documentstyle[a4wide,epsf,11pt,titlepage]{article}
\def\third{{3$^{\rm rd}$}}
\def\deg{{$^{\circ}$}}
\def\bd17{BD +17\deg 3248}
\def\cs22{CS~22892$-$052}

\newcounter{nref}
\newcommand{\bbib}{%
  \renewcommand{\refname}{\large\bf References}%
  \begin{thebibliography}{99}%
  \setcounter{enumiv}{\thenref}}
\newcommand{\ebib}{%
  \end{thebibliography}%
  \setcounter{nref}{\arabic{enumiv}}}
\newcommand{\head}[3]{%
  \setcounter{nref}{0}%
  \thispagestyle{empty}%
  \section*{\LARGE\bf #1}%
  \stepcounter{section}%
  \addcontentsline{toc}{section}{#1}%
  \large\itshape%
  #2\\\vspace{0.1pt}\\%
  #3%
  \normalsize\upshape%
  \bigskip}

\begin{document}


\head{Halo Star Abundances and Heavy Element Nucleosynthesis}
     {J. J. \ Cowan$^1$, C.\ Sneden$^2$}
     {$^1$ Department of Physics \&  Astronomy, 
University of Oklahoma, Norman, OK
73019\\
      $^2$ Department of Astronomy and McDonald Observatory, University of 
Texas, Austin, TX 78712}



\subsection{Introduction}

Abundance observations of neutron-capture elements in halo stars are 
providing a wealth of new information on the nature of 
heavy element nucleosynthesis 
in evolved  stars. In particular these observations provide clues and 
constraints to the synthesis mechanisms for the $n$-capture elements,
suggestions for the site (or sites) for the $r$-process and 
Galactic chemical evolution. 
We have been obtaining such stellar abundance data both with 
ground-based telescopes and recently with the Hubble Space Telescope
(HST). We present a small sample  of the results of our  new observations, 
along with some 
preliminary analysis,  
in this short proceedings paper. A more complete listing of the new 
HST abundance data and more detailed analyses will
be forthcoming in Cowan {\it et al.} \cite{cowan.1}.

\subsection{n-Capture Observations and the r-Process}

\begin{figure}[ht]
   \centerline{\epsfxsize=0.8\textwidth\epsffile{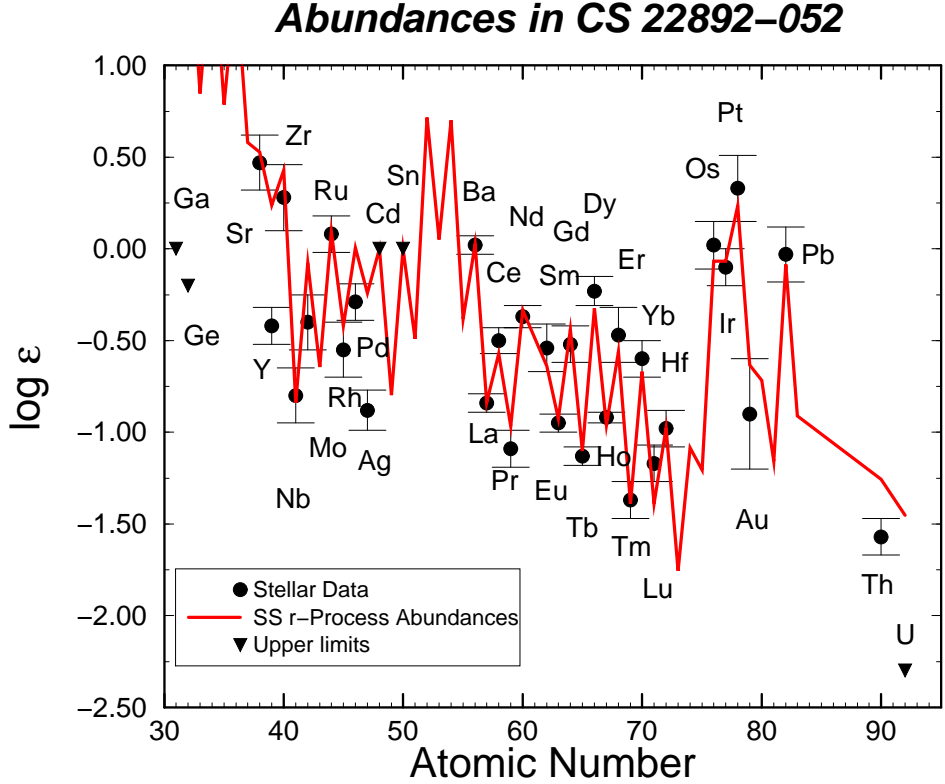}}
  \caption{$n$-capture elemental abundances in \cs22 compared to 
a (scaled for metallicity) solar system $r$-process abundance curve
(solid line).}
  \label{cowan.fig1}
\end{figure}

We show in Figure~\ref{cowan.fig1} the latest observations 
of the well-known $r$-process-rich Galactic halo star 
\cs22 \cite{cowan.2},  along with a scaled solar system
$r$-process abundance distribution (solid line). 
A total of 57 elements (52 detections and 
five upper limits) have been observed in \cs22 -- more than in any other
star except the Sun. As has been noted  previously, the agreement between
the heavier $n$-capture elements (Ba and above) and  the solar 
$r$-process abundances is quite striking \cite{cowan.3,cowan.4}. This agreement 
has been expanded and 
strengthened with our new HST observations of the \third \ $r$-process
peak elements Os and Pt in \cs22. More accurate and reliable abundance
determinations have also been obtained for the elements Nd \cite{cowan.5} 
and Ho \cite{cowan.6} as a result of new
atomic physics data.  
Ga and Ge are not observable from the ground and necessitated
the HST. While attempts to detect these elements in \cs22  were
unsuccessful,  meaningful upper limits were established
(see Figure~\ref{cowan.fig1}).
Those abundance limits fall far below the solar system $r$-process
curve and suggest that the synthesis of Ga and Ge may be different than
that (for example) of the rare-earth elements, and may be tied to the 
(very low) stellar iron abundance.

The agreement between the solar system $r$-process curve and the 
heavy (Z $\ge$ 56) elemental
abundances in \cs22 has now been seen in several other 
halo stars \cite{cowan.7}, and suggests a robust $r$-process operating over
many billions of years. It also implies a well-defined range of astrophysical
conditions ({\it e.g.}, neutron number densities) and/or that not all 
supernovae are responsible for the $r$-process -- instead perhaps only
a narrow mass range. 
We note, however,  that the well-studied 
stars showing this pattern are all $r$-process-rich and much less is know
about $r$-process-poor stars, such as HD 122563. 

What has not been as well explored in the halo stars 
is the elemental abundance
regime of Z =  40-50.
The new data for \cs22,  
along with some very sparse  data in several other
stars,  apparently indicate  a different abundance pattern between the elements 
in this regime and those for Ba and above.
Thus we see in Figure~\ref{cowan.fig1} that the elements from 
Z = 40-50,  
in general, fall below the same curve that fits the heavier (upper) end of 
the $n$-capture element distribution. This seems to suggest different
synthesis sources or origins  for the two different ranges of 
$n$-capture elements. These results strengthen earlier  
suggestions of two $r$-process sites  for the lighter and heavier 
$n$-capture elements \cite{cowan.8}. Those two sites could be different
mass ranges or frequencies of supernovae \cite{cowan.9} 
or perhaps a combination of 
supernovae and neutron-star binary mergers. Alternatively,  all of the
$n$-capture elements might be synthesized under different sets of 
conditions in the same core-collapse supernova (see Cowan and 
Sneden \cite{cowan.6} and 
references therein for further discussion). 

We note in Figure~\ref{cowan.fig1} that 
the elements Sr and Zr appear to lie on the solar $r$-process curve 
for \cs22, while Y does not. Extensive observations of these elements
versus metallicity in other stars indicates that the nucleosynthetic
origin  of Sr-Y-Zr is different than heavier elements such as Ba 
\cite{cowan.10}. Furthermore, there is a possibility 
that a primary process --
tentatively identified as a lighter element primary process          
(LEPP) \cite{cowan.10} -- might be responsible for some fraction of 
the synthesis of each of these three elements.

\subsection{Abundance Trends in Halo Stars}

\begin{figure}[ht]
   \centerline{\epsfxsize=0.5\textwidth\epsffile{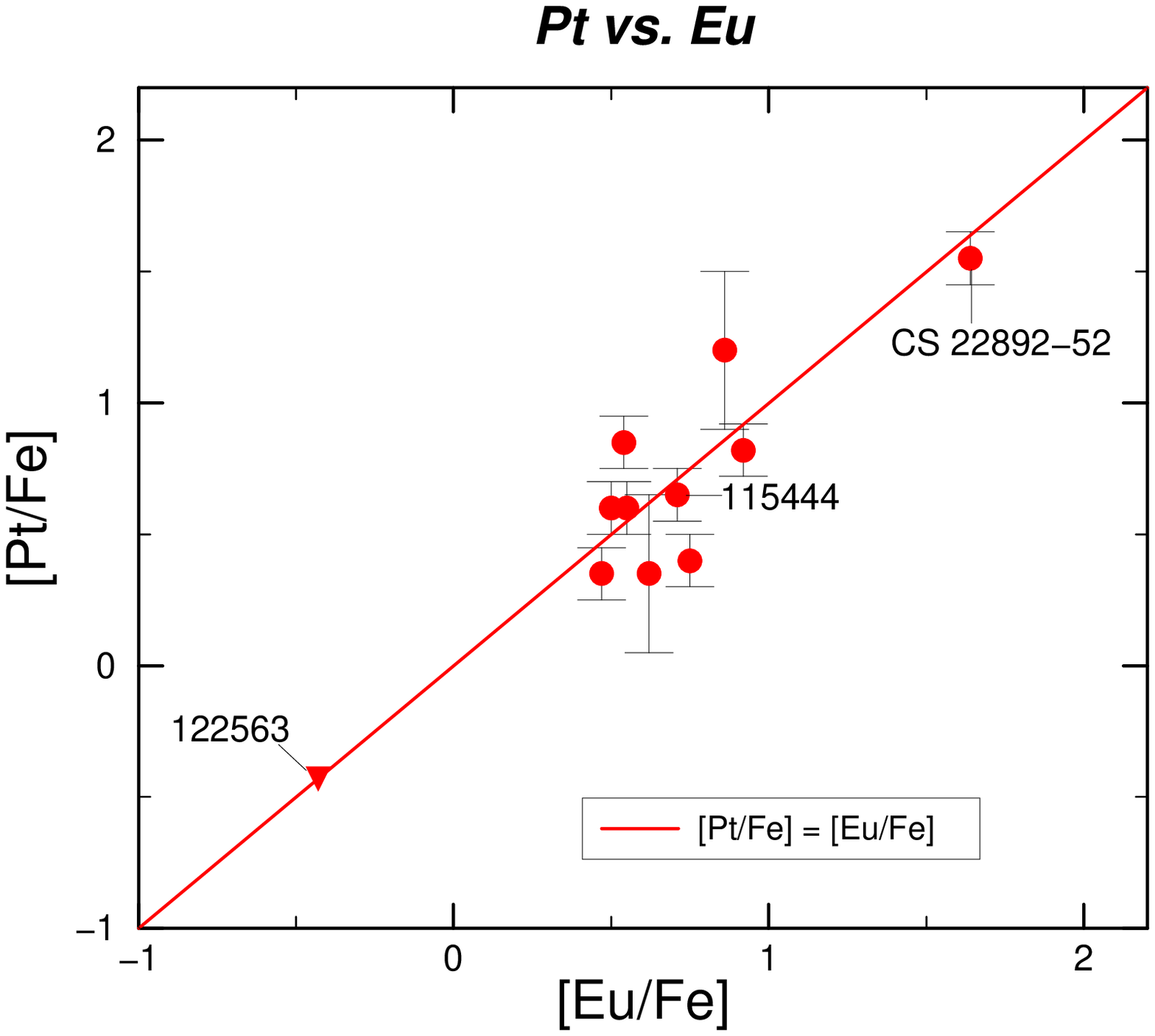}
               \epsfxsize=0.5\textwidth\epsffile{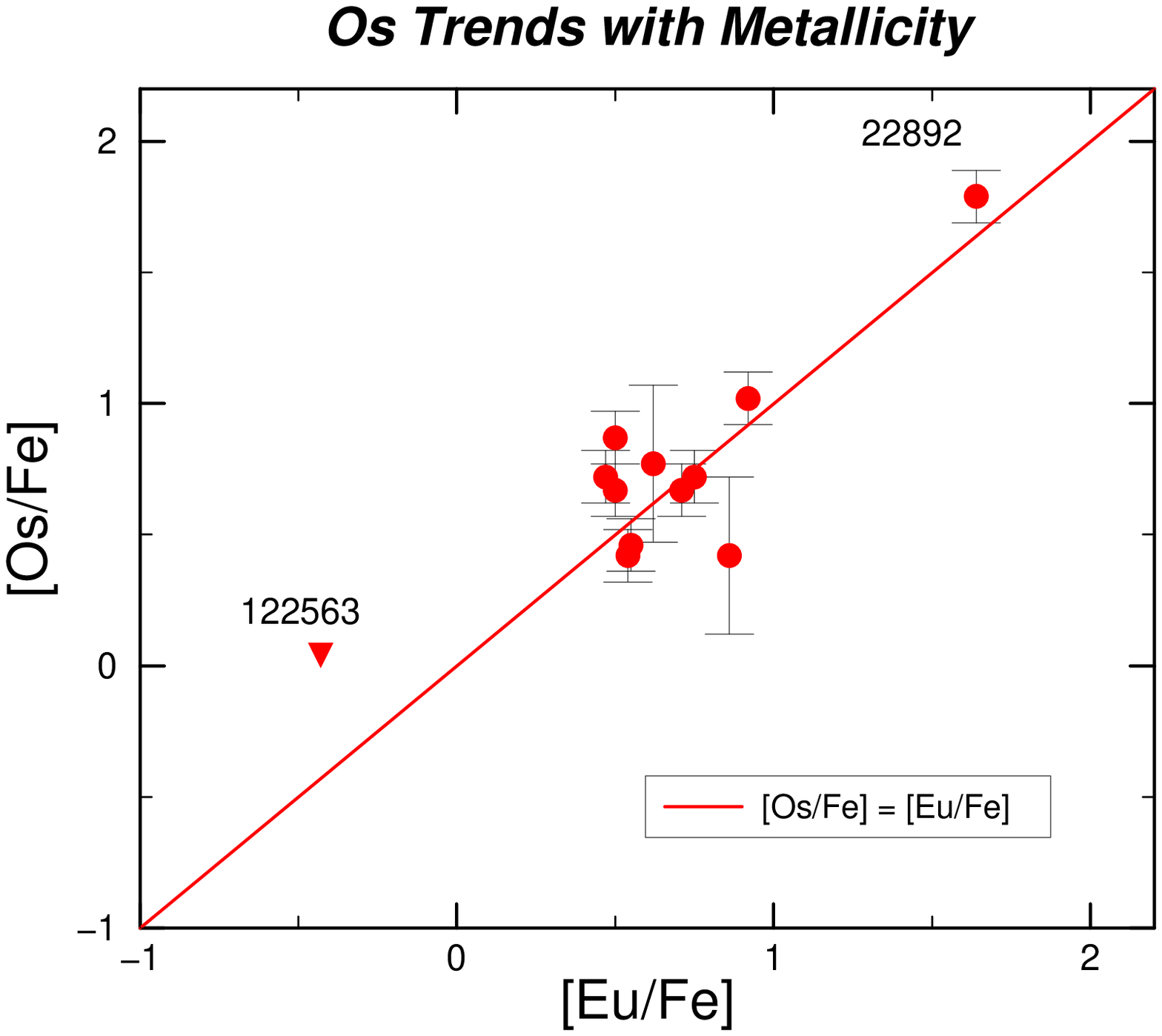}}
  \caption{Abundance correlation between [Pt/Fe] (left panel) and 
[Os/Fe] (right panel) and [Eu/Fe].}
  \label{cowan.fig2}
\end{figure}

Our new HST abundance observations of a sample of 11 halo stars have   
provided new information not 
previously attainable concerning  both the very light $n$-capture elements 
(such as Ge) and the \third \ $r$-process peak elements including 
Pt and Os. While Ge was not detected in \cs22 it was found in many of the
other target stars. Zr was also detected in these stars using HST. 
The results of those observations will be forthcoming \cite{cowan.1}. 
We show here  
the elemental abundance trends of [Pt/Fe] and [Os/Fe] 
with respect to [Eu/Fe]  in Figure~\ref{cowan.fig2}. 
The consistency between the solar system curve and the halo star abundances
(for the elements from Ba and above) 
has, in the past, mostly been predicated on the rare-earth elements
detected from the ground. Now 
with the HST detections of elements such as Os and Pt -- 
with dominant spectral transitions in the near UV -- that agreement seems 
to extend through the \third \ $r$-process peak. We see  a direct comparison
of these two latter  elements with the $r$-process element Eu -- the 
abundances of this element were obtained
with ground-based observations \cite{cowan.11}. It is clear 
in Figure~\ref{cowan.fig2} that the abundances of both Pt and Os seem
to be correlated -- there is a 45\% angle straight-line relationship -- 
with Eu in these metal-poor halo stars. 
This strongly suggests a similar synthesis origin for all three  
of those $n$-capture elements in the halo star progenitors.

\subsection*{Acknowledgments}

We thank our collaborators for their contributions to this work.
This research has been supported in part by NSF grants AST-9986974  and
AST-0307279 (JJC), AST-9987162  and AST-0307495 (CS), and by STScI grants
GO-8111 and GO-08342.


\end{document}